\documentclass[twocolumn,showpacs,preprintnumbers,nofootinbib,amsmath,amssymb,floatfix,aps,prl,10pt]{revtex4-1}
\usepackage{graphicx}
\usepackage{subfigure}
\usepackage{isotope}

\newcommand{\genie}{\textsc{genie}}

\newcommand{\globes}{\textsc{gl}{\small o}\textsc{bes}}
\graphicspath{{figs/}}

\begin{document}

\title{Nuclear effects are relevant to the calorimetric reconstruction of neutrino energy}
\author{Artur M. Ankowski}
\email{ankowski@vt.edu}
\affiliation{Center for Neutrino Physics, Virginia Tech, Blacksburg, Virginia 24061, USA}

\date{\today}%

\begin{abstract}
As the calorimetric method of neutrino-energy reconstruction is generally considered to be largely insensitive to nuclear effects, its application seems to be an effective way for reducing systematic uncertainties in oscillation experiments.
To verify the validity of this opinion, we quantitatively study the sensitivity of the calorimetric energy reconstruction to the effect of final-state interactions in an ideal detector and in a realistic scenario. We find that when particles escaping detection carry away a non-negligible fraction of neutrino energy, the calorimetric reconstruction method becomes sensitive to nuclear effects which, in turn, affects the outcome of the oscillation analysis. These findings suggest that the best strategy for reduction of systematic uncertainties in future neutrino-oscillation studies---such as the Deep Underground Neutrino Experiment---is to increase their sensitivity to particles of low energy. The ambitious precision goals appear also to require an extensive development of theoretical models capable of providing an accurate predictions for exclusive cross sections of well-controlled uncertainties.
\end{abstract}



\maketitle

In measurements of neutrino oscillations, the oscillation parameters---mixing angles, mass splittings, and the Dirac phase---are extracted from the energy dependence of collected event distributions. An accurate determination of the parameter values, therefore, requires an accurate determination of neutrino-energy spectrum. As the next generation of neutrino experiments are expected to determine the mass splittings with precision exceeding 1\%~\cite{Abe:2015zbg,Acciarri:2015uup}, they need to be able determine neutrino energies with precision exceeding 1\%. This stringent requirement translates into keeping energy uncertainties below 5 and 25 MeV, depending on the considered setup.

Having at their disposal only polychromatic beams, oscillation experiments need to reconstruct neutrino energies on an event-by-event basis from the measured kinematics of particles produced in interactions between the beam and the detector target.

At energies exceeding several hundred MeV, where different dynamical mechanisms contribute to the cross section~\cite{Formaggio:2013kya}, the calorimetric method of neutrino-energy reconstruction~\cite{Michael:2008bc,Ayres:2007tu,Acciarri:2016crz} is of particular importance because of its universal applicability. In this method, the neutrino energy is estimated by adding the energies deposited in the detector by all observed interaction products.

Recently~\cite{Ankowski:2015jya,Ankowski:2015kya}, it has been shown that finite detection capabilities---energy resolutions, efficiencies and thresholds---may have a sizable affect on the outcome of such energy reconstruction and on the extracted oscillation parameters. These findings point toward the relevance of an accurate determination of the detector response in extensive exposures to a variety of test beams in neutrino-oscillation experiments employing the calorimetric method of energy reconstruction.

As in Refs.~\cite{Ankowski:2015jya,Ankowski:2015kya} uncertainties related to the nuclear model has not been considered, here we discuss the sensitivity of the calorimetric reconstruction to nuclear effects. For the purpose of this complementary study, we assume that the detector response to all particles is perfectly known and accurately accounted for in the simulations used in the oscillation analysis, and we only investigate if the calorimetric reconstruction exhibits a sensitivity to nuclear effects. To this end, we analyze the role of final-state interactions (FSI) between the products of the primary neutrino interaction in the target nucleus and the residual nuclear system.

While the kinematic method of neutrino energy reconstruction is well-known to be affected by nuclear effects~\cite{Benhar:2009wi,Leitner:2010kp,Martini:2012fa,Nieves:2012yz,Meloni:2012fq,Lalakulich:2012hs,Martini:2012uc,Coloma:2013rqa, Coloma:2013tba,Mosel:2013fxa,Jen:2014aja,Ankowski:2014yfa,Ericson:2016yjn,Ankowski:2016bji}, this is not the case for the calorimetric method. Although FSI effects can redistribute the energy transferred to the nucleus between a number of particles, they do not affect the total energy~\cite{Mosel:2015yaa}. As a consequence, one would obtain the true value of the neutrino energy for any charged-current event by summing the kinetic energies of knocked-out nucleons, the total energies of other (charged and neutral) particles produced in the interaction, and both the recoil and (possibly delayed) deexcitation energy of the residual nucleus, provided that detector effects play a negligible role.

However, when detector effects are relevant, the neutrino energy reconstructed calorimetrically becomes sensitive to the event composition, because detector effects, in general, differ for different particle species and exhibit an energy dependence. For example, the outcome of the reconstruction is going to be different if strong FSI effects redistribute the struck-nucleon's energy between a few low-energy nucleons below the detection threshold, or if they lead to a significant energy transfer from a~proton to a neutron which, in turn, escapes detection.

This article addresses the question of whether an accurate description of nuclear effects is relevant for the calorimetric method of neutrino-energy reconstruction, discussing the influence of FSI effects on the outcome of a $\nu_\mu$ disappearance analysis. We treat systematic uncertainties, implement the $\chi^2$ calculations, and determine the confidence regions following Refs.~\cite{Coloma:2012ji,Coloma:2013rqa,Coloma:2013tba}. Employing the software package \globes{}~\cite{Huber:2004ka,Huber:2007ji,Coloma:2012ji}, we consider an idealized experiment with a narrow-band off-axis beam peaked at energy $\sim$600 MeV~\cite{Huber:2009cw} that collects data for 5 years at the beam intensity 750 kW. The near (far) detector of fiducial mass 1.0 kt (22.5 kt) is located at 1.0 km (295.0 km) from the neutrino source. The detectors are assumed to be functionally identical.

\begin{figure*}
\begin{center}
\includegraphics[width=0.42\textwidth, viewport = 96 282 454 534]{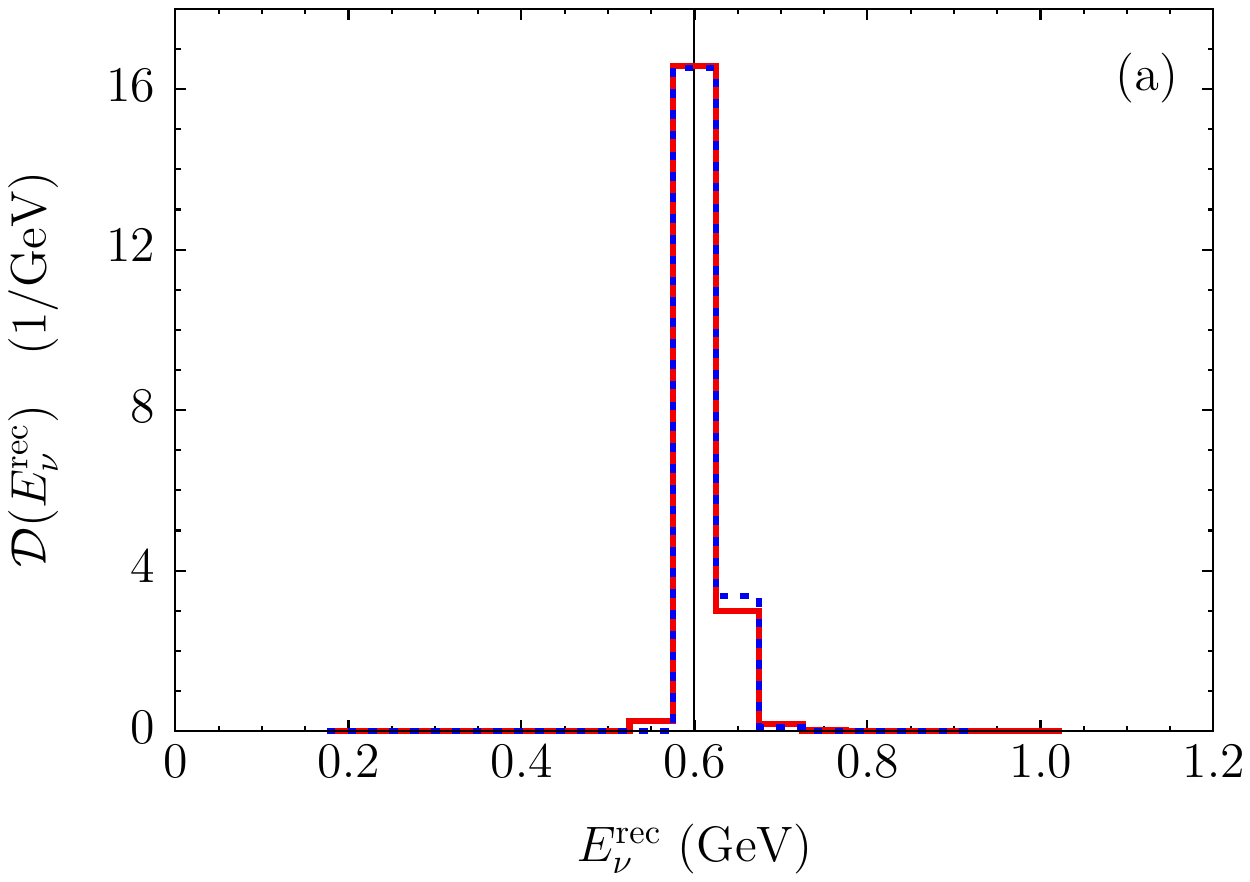}
\hspace{0.8cm}
\includegraphics[width=0.42\textwidth, viewport = 96 282 454 534]{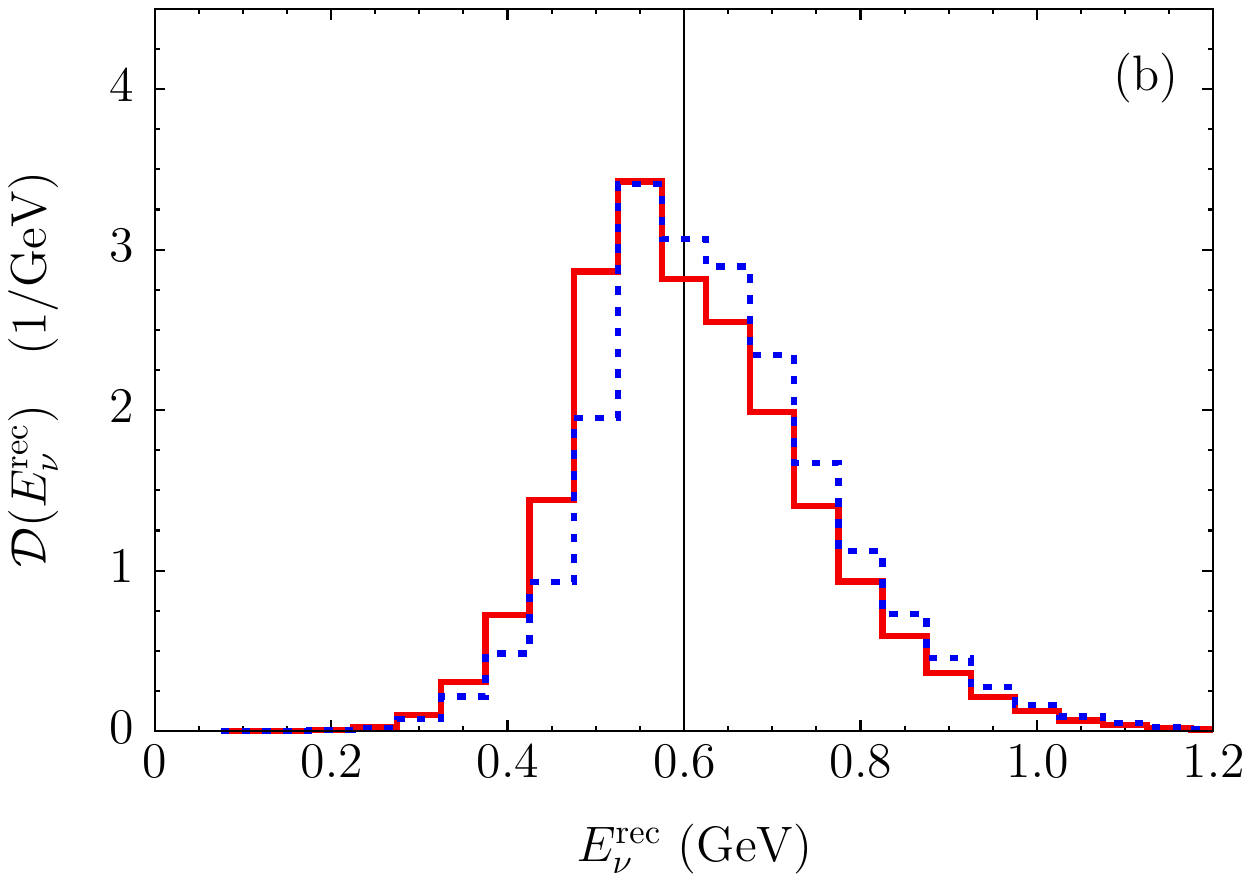}
    \subfigure{\label{fig:eRec_noDE}}
    \subfigure{\label{fig:eRec_DE}}
\caption{\label{fig:eRec} (color online). Sensitivity to nuclear effects of the distributions of neutrino energy reconstructed calorimetrically in the (a) absence and (b) presence of detector effects, calculated for quasielastic scattering with any number of knocked-out nucleons at the true neutrino energy 0.6 GeV. The solid and dotted lines represent the results accounting for and neglecting the effect of final-state interactions, respectively. The corresponding differences in the mean values of the reconstructed energy are (a) $\sim$1 MeV and (a) $\sim$25 MeV. Note that the histograms in panel (a) largely overlap and that the vertical scales in the panels differ.}
\end{center}
\end{figure*}

To minimize uncertainties of the description of nuclear effects, playing central role in this analysis, we analyze carbon as the detector material, as in the previous works~\cite{Ankowski:2015jya,Ankowski:2015kya,Ankowski:2016bji}. For this target, an overwhelming majority of the available data for neutrino cross sections have been extracted~\cite{Ankowski:2016jdd} and nuclear models implemented in Monte Carlo generators can be expected to involve the lowest uncertainties.

We emphasize that while the considered experimental setup is similar to that of T2K~\cite{Abe:2011ks}, it is highly idealized and differs in the selection of the target material, the energy reconstruction method, and the procedure of the oscillation analysis.

\begin{table}[b]
\caption{\label{tab:resolutions}Parameters of our simulation of detector effects, comparable to those of the DUNE's Fast Monte Carlo~\cite{Alion:2016uaj}.}
\begin{ruledtabular}
\begin{tabular}{l l l r}
Particle type & $a_h$ & $b_h$ & $t_\text{thres}$ (MeV)\\
\hline
$\pi^\pm$ & 0.00 & 0.05 & 100\\
$\pi^0$  & 0.00  & 0.30 & 100\\
$e$, $\gamma$    & 0.15 &  0.02 &  30\\
$\mu$  & 0.00 &  0.02 &  30\\
$n$      & 0.40 & 0.00  &  50\\
$p$, other      & 0.30 &  0.05 &  50\\
\end{tabular}
\end{ruledtabular}
\end{table}

In our study the following mechanisms of neutrino interactions are considered: quasielastic scattering with (i) one or (ii) more nucleons in the final state, (iii) resonant pion production, and (iv) deep-inelastic scattering. We obtain the migration matrices using the Monte Carlo generator \genie{} 2.8.0~\cite{Andreopoulos:2009rq} supplemented with the $\nu T$ package of additional modules~\cite{Jen:2014aja}. This choice allows us to describe the ground state properties of the carbon nucleus using the realistic spectral function~\cite{Benhar:1989aw,Benhar:1994hw} in the dominant quasielastic channel of neutrino interactions.

Accounting for detector effects, we assume perfect detection efficiency for every particle of the true kinetic energy exceeding its detection threshold $t_\text{thres}$. To simulate the effect of finite energy resolutions, we smear the kinetic energies of hadrons, $T_h$, according to the Gaussian distributions of the width $\sigma(T_h)$,
\[
\frac{\sigma(T_h)}{T_h}=\sqrt{\frac{a_h^2}{T_h/\textrm{GeV}}+b_h^2}.
\]
In the case of charged leptons, we smear their momenta instead of the kinetic energies. The values of the parameters $a_h$ and $b_h$ employed in this analysis are given in Table~\ref{tab:resolutions}. Note that they are comparable to those used in the Fast Monte Carlo of the Deep Underground Neutrino Experiment (DUNE)~\cite{Alion:2016uaj}.

\begin{figure*}
\begin{center}
\includegraphics[width=0.42\textwidth]{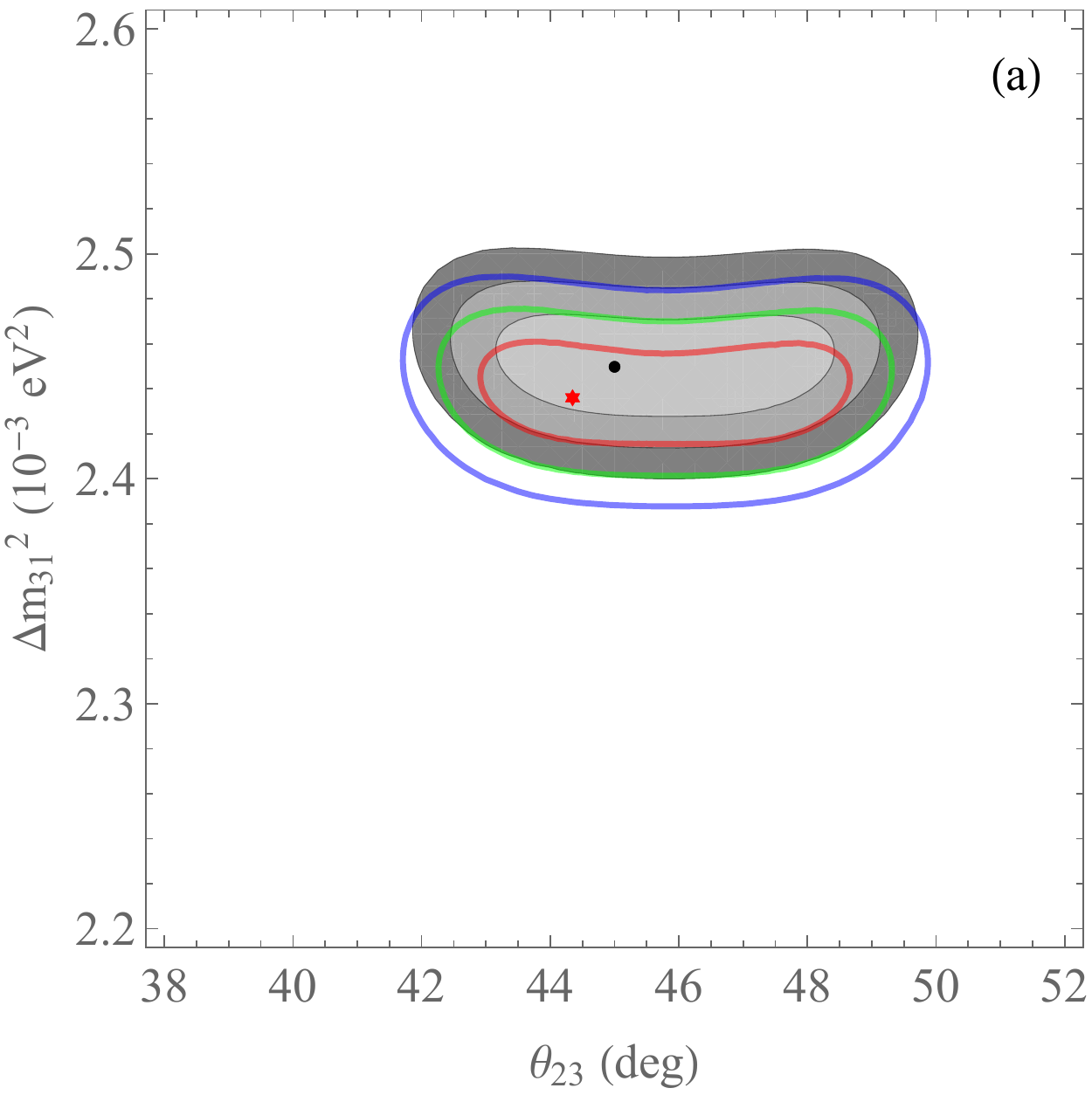}
\hspace{0.8cm}
\includegraphics[width=0.42\textwidth]{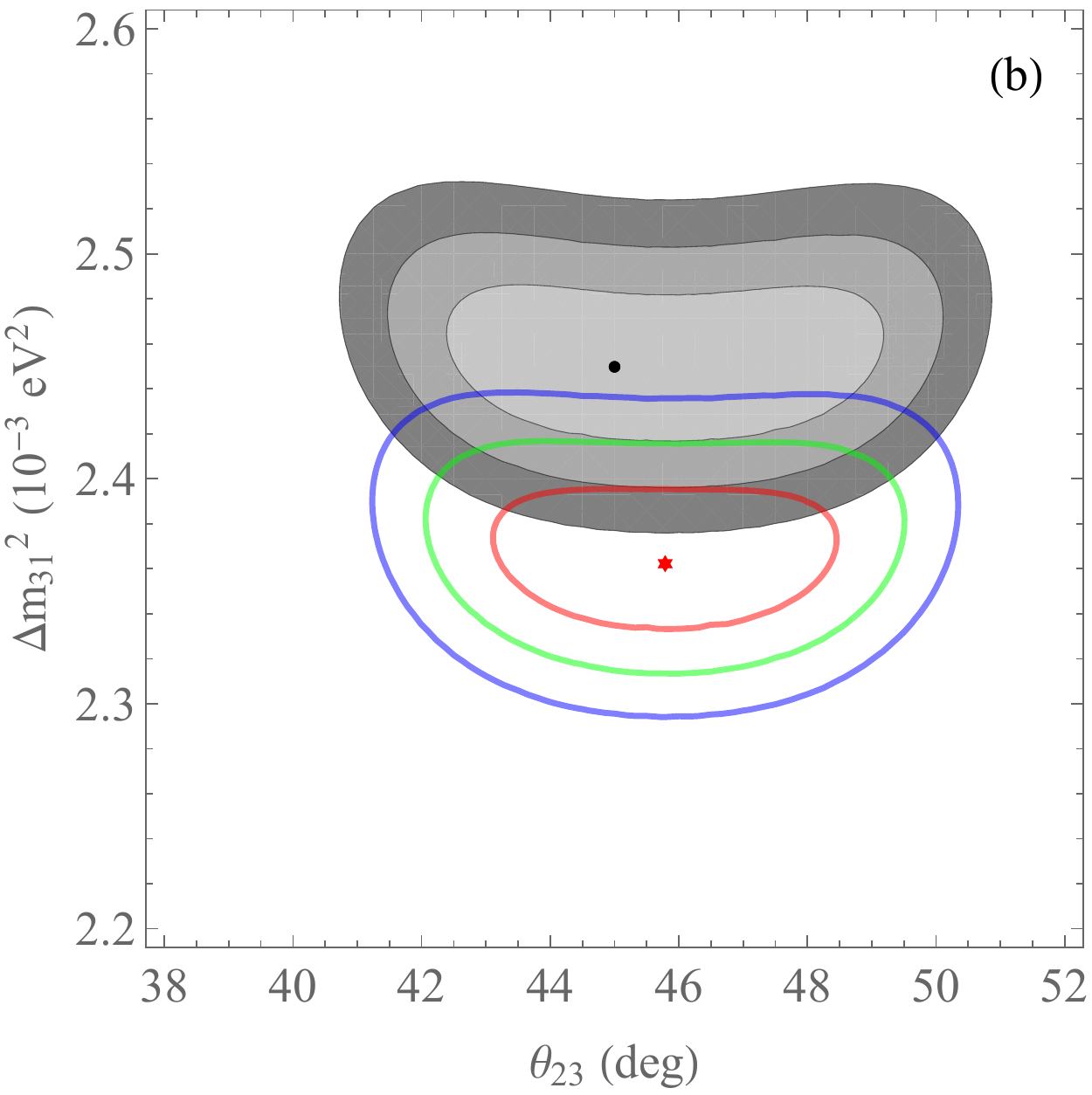}
    \subfigure{\label{fig:oscill_noDE}}
    \subfigure{\label{fig:oscill_DE}}
\caption{\label{fig:oscill} (color online). Sensitivity to nuclear effects of the oscillation analysis employing the calorimetric method of neutrino-energy reconstruction in the (a) absence and (b) presence of detector effects. The 1, 2, and 3$\sigma$ confidence regions in the $(\theta_{23},\,\Delta m_{31}^2)$ plane are obtained using the migration matrices accounting for (shaded areas) or neglecting (solid lines) the effect of final-state interactions, included in the simulated data. The star and circle show the best fit point and true values of the oscillation parameters, respectively.}
\end{center}
\end{figure*}

Being in nuclear medium, the nucleon struck in neutrino scattering off the nucleus interacts with the spectator nucleons. When these interactions are ``soft'', they can be accounted for as a modification of the struck-nucleon's energy spectrum. Note that modifying the allowed struck-nucleon's energies, this effect modifies also the charged-lepton's energy. In the (dominant) forward kinematics, it decreases the energy transfer to the nucleus and lowers the kinetic energies of particles escaping from the nucleus. Its role is reversed when the charged lepton is produced at high angle~\cite{Ankowski:2014yfa}.

On the other hand, ``hard'' interactions altering the struck-nucleon's momentum lead to a development of an intranuclear cascade of particles, as a consequence of which multiple particles leave the residual nucleus and the kinetic energies of the primary interaction products are typically lowered. This effect does not modify the charged-lepton's energy.

Here we consider ``hard'' secondary interactions for all neutrino-scattering mechanisms, relying on the cascade implemented in \genie{} 2.8.0~\cite{Andreopoulos:2009rq}. In the quasielastic channel, we also include ``soft'' reinteractions, describing them by means of the real part of the optical potential~\cite{Ankowski:2014yfa,Cooper:1993nx}.

As shown in Fig.~\ref{fig:eRec_noDE} for quasielastic $\nu_\mu$ scattering with any number of knocked-out nucleons, the calorimetric method of neutrino-energy reconstruction is not affected by nuclear effects, such as FSI, when detector effects play negligible role. The reconstructed energy distributions---elements of the migration matrices---obtained with and without accounting for FSI effects largely overlap and their mean values differ by as little as $\sim$1 MeV at the true energy 600 MeV.

This is, however, not the case when detector effects are important. As the particles of kinetic energies below their detection thresholds carry away the energy from the event, they shift the reconstructed energy toward lower values. FSI effects enhance this shift by redistributing the transferred energy between multiple particles and lowering their kinetic energies, which increases the undetected energy. This feature is illustrated in Fig.~\ref{fig:eRec_DE}, showing that due to detector effects there is a visible difference between the reconstructed energy distributions calculated with and without FSI effects, corresponding a $\sim$25 MeV discrepancy between the mean values of the reconstructed energy. Reducing the energy transfer to the nucleus, the real part of the optical potential effectively increases the energy resolution, as can be observed at the reconstructed energy $E^\text{rec}_\nu\sim0.55$ GeV in Fig.~\ref{fig:eRec_DE}. This is a consequence of much higher energy resolution for muons than for hadrons, see Table~\ref{tab:resolutions}. We note, however, that this effect turns out to be small for the carbon target compared to the influence of intranuclear cascade.

The observed sensitivity of the $E^\text{rec}$ distributions to FSI effects---induced by detector effects---has important implications for the oscillation studies. Should the fraction of neutrino energy carried away by undetected particles be negligible, experiments using the calorimetric energy reconstruction would be largely insensitive to inaccuracies of the nuclear model employed in the oscillation analysis. We illustrate it in Fig.~\ref{fig:oscill_noDE}, presenting the 1, 2 and 3$\sigma$ confidence regions in the $(\theta_{23}, \Delta m^2_{31})$ plane for the true event rates simulated accounting for FSI effects. The shaded areas represent the results for the fitted rates obtained accounting for FSI effects. The solid lines correspond to the fitted rates calculated neglecting FSI effects. While there are small differences between the extracted and true values of the oscillation parameters, they can be attributed to imperfections of the energy conservation in the employed intranuclear cascade. Those imperfections are visible in Fig.~\ref{fig:eRec_noDE}, where the reconstructed energy differs from the true one, $E_\nu=0.6$ GeV, by more than 25 MeV in $\sim$17\% of events.
The results of Fig.~\ref{fig:oscill_noDE} show that even neglecting FSI effects altogether does not affect the extracted oscillation parameters significantly, provided that detector effects do not play important role.

On the other hand, when particles escaping detection have a non-negligible contribution to the event energy, the outcome of the oscillation analysis exhibits strong dependence on FSI effects. Figure~\ref{fig:oscill_DE} shows how the confidence regions from Fig.~\ref{fig:oscill_noDE} change when detector effects are included in the analysis. It clearly appears that in this case, FSI effects cannot be neglected and that the calorimetric method of energy reconstruction is not insensitive to the nuclear model employed in the data simulations.

These results do not came as a surprise. In a realistic situation, a nuclear model is necessary to correct the energy deposited in the detector for the missing energy---carried away by undetected particles---in order to estimate the neutrino energy accurately. Any inaccuracies in the description of nuclear effect may translate into a bias in the reconstructed energy which, in turn, leads to a bias in the extracted values of the oscillation parameters. Obtaining the confidence levels represented by lines in Fig.~\ref{fig:oscill}, we have neglected FSI effects in the calculations of the fitted rates. In this way, we have misreconstructed the neutrino energy, which is known to affect predominantly the extracted mass splitting.

Obviously, in the oscillation analysis of a real experiment, FSI effects are always accounted for. However, our results point out the importance of an \emph{accurate} description of nuclear effects---including those related to FSI---when the calorimetric method is employed to reconstruct neutrino energy. As the presented results are obtained for the carbon target, the observed effect can be expected to be even more significant in the case of experiments using argon as the detector material, due to its higher mass number and to the difference between the proton and neutron numbers. Moreover, as the multiplicity of particles produced in the interaction is an increasing function of neutrino energy, FSI effects can also be expected to be more significant at the kinematics of DUNE---dominated by pion production---then at the energies considered here.

The best way to mitigate the sensitivity to nuclear effects in future neutrino-oscillation experiments is to reduce the missing energy by lowering detection thresholds and increasing detection efficiencies. As in the presented analysis we assume that all particles of energies above their thresholds are measured, the amount of missing energy is driven by the threshold values alone. We observe that if the proton and neutron thresholds were lowered to 25 MeV (30 MeV), the difference between the extracted and true values of the oscillation parameters presented in Fig.~\ref{fig:oscill_DE} would be within 1$\sigma$ (2$\sigma$) confidence region. This is a consequence of the dominant contribution of the nucleon energy to the hadronic energy of events at the considered kinematics.

On the other hand, lowering the pion detection thresholds can be expected to be of particular importance for DUNE, at the kinematics of which pion production is the main mechanism of neutrino interactions. Should these thresholds be $100$ MeV, as assumed in the DUNE's Fast Monte Carlo~\cite{Alion:2016uaj}, a single undetected pion would result in an underestimation of the reconstructed neutrino energy in the event by up to $t_\text{thres} + m_\pi\simeq 240$ MeV, $m_\pi$ being the pion mass.
In order to bring DUNE closer to the assumed precision of the oscillation analysis exceeding 1\%---corresponding to the energy-reconstruction accuracy better than 25 MeV---it would be beneficial to establish the sensitivity to 4.1-MeV muons produced in $\pi^\pm$ decay at rest (lifetime 26.0 ns) or to Michel electrons of energies below 52.8 MeV from subsequent muon decay at rest (lifetime 2.20 $\mu$s).

Owing to the required precision of neutrino-energy determination and the difficulty with an accurate measurement of the neutron contribution to the event energy, it seems that the ambitious program of the next generation of oscillation experiments is going to rely heavily on theoretical models employed in Monte Carlo simulations which, in turn, demands their urgent development and extensive tests of the predicted hadron spectra against experimental data. An essential ingredient of such models is an accurate description of the ground-state properties of the nuclear targets of interest, in addition to an accurate modeling of FSI effects.

In summary, we have discussed the sensitivity of calorimetric energy reconstruction to nuclear effects, analyzing the effect of final-state interactions. We have found that they play important role when particles escaping detection carry away a non-negligible fraction of neutrino energy. In view of these observations, the best strategy for future neutrino-oscillation experiments seems be to minimize detection thresholds in their detectors and to perform extensive validations of the accuracy of nuclear models employed in data analysis. Moreover, development of theoretical models capable of providing an accurate predictions for exclusive cross sections appears to be a prerequisite for the success of precise oscillation measurements.

\begin{acknowledgments}
I am deeply indebted to Pilar Coloma for her invaluable help at all stages of this project and for her hospitality during my stay at Fermilab. Special thanks are addressed to John Beacom, Omar Benhar, Patrick Huber, Jonathan Link, and Camillo Mariani for their comments on the manuscript. This work has been supported by the Visiting Scholars Award Program of the Universities Research Association under Award No. 16-F-1 and by the National Science Foundation under Grant No. PHY-1352106.
\end{acknowledgments}


%

\end{document}